\documentclass[prd,aps,showpacs,preprintnumbers,amsmath,amssymb]{revtex4}
\usepackage{amsmath,amssymb,graphicx}

\def\be{\begin{equation}}
\def\ee{\end{equation}}
\def\bea{\begin{eqnarray}}
\def\eea{\end{eqnarray}}

\newcommand{\nc}{\newcommand}
\nc{\ba}{\begin{eqnarray}}
\nc{\ea}{\end{eqnarray}}

\newcommand\s{\sigma}

\newcommand\la{\lambda}
\newcommand\ti{\tilde}
\nc{\ga}{\gamma}
\nc{\x}{{\bf x }}
\nc{\kk}{{\bf k }}
\nc{\f}{{\bf f }}
\nc{\e}{{\bf e }}
\nc{\T}{ \theta (s_i (t)- \s) }
\nc{\TT}{ \theta (s_i (t_{ r \, i } )- \s) }
\nc{\br}{   (s_i (t)- \s)  }
\nc{\tphi}{\tilde{\phi}}
\nc{\tA}{\tilde{\cal{A}}}

\input{epsf.sty}

\begin{document}

%%%%%%%%%%%%%%%%%%%%%%%%%%%%%%%%%%%%%%%%%%%%%%%%%%%%%%%%%%%%%%%%%%%%%%%
\title{A Radiation Bounce from the Lee-Wick Construction?}

\author{Johanna Karouby and Robert Brandenberger}

\affiliation{Department of Physics, McGill University, Montr\'eal, QC, H3A 2T8, Canada}


\begin{abstract}

It was recently realized that matter modeled by the scalar field sector of the Lee-Wick
Standard Model yields, in the context of a homogeneous and isotropic cosmological
background, a bouncing cosmology. However, bouncing cosmologies induced by
pressure-less matter are in general unstable to the addition of relativistic matter (i.e. radiation).
Here we study the possibility of obtaining a bouncing cosmology if we add not only
radiation, but also its Lee-Wick partner, to the matter sector. We find that, in general,
no bounce occurs. The only way to obtain a bounce is to choose initial conditions
with very special phases of the radiation field and its Lee-Wick partner.
  
\end{abstract}

\maketitle

\section{Introduction}

The inflationary scenario \cite{Guth} is the current paradigm of early universe cosmology.
It addresses some of the problems which the previous paradigm, the Standard Big Bang
model, could not address, and it gave rise to the first theory of cosmological
structure formation based on fundamental physics \cite{Mukh} whose predictions
were later confirmed by the precision observations of the cosmic microwave background.
Inflationary models, however, are faced with serious conceptual problems (see e.g.
\cite{RHBrev}), among which the singularity problem and the ``Trans-Planckian"
problem for fluctuations. In the context of General Relativity as the theory of space-time,
it has been shown \cite{Borde} that inflationary models have a singularity in the past and therefore
cannot yield a complete theory of the early universe. The ``Trans-Planckian" problem
for fluctuations \cite{RHBrev,Jerome} relates to the fact that in inflationary models,
the wavelengths of perturbation modes which are observed today were smaller
than the Planck scale in the early periods of inflation, and were thus in the 
``short wavelength zone of ignorance" in which we cannot trust the theory which is
being used to track the fluctuations. In fact, in \cite{Jerome} it is shown that
the predictions for observations are in fact rather sensitive to the physics assumed
in this zone of ignorance. These conceptual problems of inflationary cosmology
form one of the motivations for considering possible alternatives to inflation.

One of the alternative scenarios to inflation is the ``matter bounce" paradigm (see e.g.
\cite{RHBcospa,RHBrio} for introductory expositions). In this scenario it is
assumed that the universe undergoes a non-singular cosmological bounce.
Time runs from $- \infty$ to $+ \infty$. The time coordinate can always be adjusted
such that the bounce point is at time $t = 0$. A space-time sketch of a bouncing
universe is given in Fig. 1. In this figure, the horizontal axis is space and the vertical
axis is time. We sketch the Hubble radius $H(t)^{-1}$, the scale which separates
wavelengths on which microphysics dominates (sub-Hubble) from those where matter
forces are frozen out (super-Hubble). If the contracting and expanding phases far
away from the bounce point are described by General Relativity and we consider
matter with pressure density $p > - \rho / 3$ (where $\rho$ is the energy
density), then it follows that scales which are currently observed exited the
Hubble radius at some point during the contracting phase. As was realized 
in \cite{Wands,FB,Wands2}, if the curvature fluctuations start out early in the
contracting phase on sub-Hubble scales in their vacuum state, then the
growth of the perturbations on super-Hubble scales during the period
of contraction leads to a scale-invariant spectrum of curvature fluctuations
on super-Hubble scales before the bounce. Detailed analyses of the
evolution of cosmological fluctuations through the non-singular bounce
performed in the context of specific bouncing models 
(see e.g. \cite{BFS,ABB,Cai1}) show that the spectrum of curvature
fluctuations is unchanged during the bounce on wavelengths which
are large compared to the bounce time, a result which agrees with what
is obtained by applying the Hwang-Vishniac matching conditions \cite{HV,DM}
to connect perturbations across a space-like ``matching" hypersurface
between a contracting and an expanding Friedmann universe.

By construction, a bouncing cosmology is non-singular.
In such a model, the wavelength of fluctuations which are being probed in 
current observations always remains far larger than typical microphysical
scales. If the energy density at the bounce point is set by the scale of
particle physics Grand Unification, then the physical wavelength corresponding
to the current Hubble radius is about $1 {\rm mm}$, to quote just one number.
Hence, the fluctuations remain in the regime controlled by the infrared limit
of the theory, far from the trans-Planckian zone of ignorance.

The challenge is to obtain a bouncing cosmology. One must either give up
General Relativity as the theory of space-time, or else one must invoke a new
form of matter which violates some of the ``usual" energy conditions
(see \cite{Hawking} for a discussion of the assumptions underlying the
singularity theorems of General Relativity). For a recent review on how
bouncing cosmologies can be obtained see \cite{Novello}. We here 
mention but a few recent attempts. Introducing higher derivative gravity
terms can lead to non-singular cosmologies, as in the ``non-singular universe
construction" of \cite{BMS}. Similarly, the ghost-free higher derivative action
of \cite{Biswas} leads to a bouncing cosmological background. Horava-Lifshitz
gravity also leads to a bouncing cosmology provided that the spatial
curvature does not vanish \cite{HLbounce}. Bouncing
cosmologies may also arise from quantum gravity, as e.g. in loop quantum
cosmology (see e.g. \cite{LQC} for a recent review). If we maintain General
Relativity as the theory of space and time, then one can obtain a bounce by
introducing new forms of matter such as ``quintom" matter \cite{quintom}. In
this case, in addition to the matter sector with regular sign kinetic action, there
is a new sector (a ``ghost" sector) which has an opposite sign kinetic action.

Several decades ago, Lee and Wick \cite{LeeWick} introduced a field
theory construction which involves degrees of freedom with opposite
sign kinetic terms. The Lee-Wick model aims at stabilizing the Higgs mass
against quadratically divergent terms and is interesting to particle physicists
since it can address the ``hierarchy problem". The Lee-Wick construction
was recently resurrected and extended to yield a ``Lee-Wick Standard Model"
\cite{Grinstein:2007mp,Wise:2009mi}. The Lee-Wick model can thus potentially
provide a framework for obtaining a bouncing cosmology. In \cite{Cai2}, the
Higgs sector of the Lee-Wick Standard Model was analyzed and it was shown
that, indeed, a bouncing cosmology emerges. However, the scalar field
Lee-Wick bounce is unstable against the addition of regular radiation to the
matter sector (as will be explained in Section 2 of this paper). Since we know
that there is radiation in the universe, one may than worry whether the Lee-Wick
bounce can be realized at all. However, to be consistent with the philosophy
of the Lee-Wick construction, Lee-Wick radiation terms with opposite sign
kinetic actions must be added. In this paper we address the question whether, in
this context, a cosmological bounce can be achieved. We find that unless
the phases of the two fields are chosen in a very special way then no
bounce will occur.
 
The outline of this paper is as follows: in the next section we briefly review the
philosophy behind the Lee-Wick model  and discuss why the scalar sector
of the Lee-Wick model taken alone would yield a bouncing cosmology.
In Section 3 we introduce the Lagrangian for Lee-Wick electromagnetism
and derive the expression for the energy density. In order to study the
cosmological implications of our action, we need to know how plane waves
of the Lee-Wick partner of the radiation field evolves. This is the focus of
Section 4. After understanding how regular and Lee-Wick radiation evolve,
we can then study under which conditions a bouncing cosmology might
result.

\section{Review of the Lee-Wick Model and the Scalar Lee-Wick Bounce}

We will review the Lee-Wick model and the Lee-Wick bounce in the simple
case of a single scalar field $\hat{\phi}$. The hypothesis of Lee and Wick
\cite{LeeWick} was to add an extra scalar degree of freedom designed to
cancel the quadratic divergences in scattering matrix elements. Originally,
the new degree of freedom was introduced by adding a higher derivative term
of the form $(\partial^2 \hat{\phi})^2$ to the action, yielding a higher order
differential equation and hence a new degree of freedom. It is, however, simpler
to isolate the new degree of freedom by introducing an auxiliary scalar field 
$\tilde{\phi}$ and redefining the ``physical" field to be $\phi$ 
(see \cite{Grinstein:2007mp}). After doing this and after a field
rotation the Lagrangian becomes
\be
{\cal L}  \, = \,  {1 \over 2} \partial_{\mu} \phi \partial^{\mu} \phi 
- \frac{1}{2} \partial_{\mu} \tphi \partial^{\mu} \tphi + \frac{1}{2} M^2 \tphi^2 
- \frac{1}{2} m^2 (\phi - \tphi)^2  - V(\phi - \tphi) \, ,
\ee
where $M \gg m$ is the mass scale of the new degree of freedom, and $V$ is the original
potential which after the field redefinition depends on both fields. 

The field $\tphi$ is called the Lee-Wick partner of $\phi$. It has the opposite
sign kinetic Lagrangian and the opposite sign of the mass square term.
Hence, without any coupling to other fields or to gravity the evolution of
$\tphi$ would be stable and consist of oscillations about $\tphi = 0$.
However, in the presence of any coupling of $\tphi$ with other fields
there are serious potential instability and unitarity problems \cite{BG,Nak,Gleeson,Cline}.
Ways to make the theory consistent were discussed many years ago in \cite{Cut}
and more recently in \cite{Tonder} in the case of interest in the current paper, namely
Lee-Wick electromagnetism.  In \cite{Tonder}, a proposal for a ultraviolet (UV) complete
theory of Quantum Electrodynamics via the Lee-Wick construction was made. It was
argued that the presence of ghost poles in virtual state propagators and the loss of 
microcausality do not necessary mean that causality is violated at macroscopic scales. 
This would be the case if the Lee-Wick particles decayed fast enough \cite{Fornal:2009xc}.

The Lee-Wick model has been resurrected in \cite{Grinstein:2007mp} with the goal
of studying signatures of this alternative model to supersymmetry in LHC experiments.
For some projects to try to test experimentally the predictions of the
Lee-Wick model see e.g. \cite{Alvarez:2009af,Chivukula:2010nw}.

Let us now review \cite{Cai2} how a non-singular bouncing cosmology can emerge
from the scalar sector of a Lee-Wick model. In fact, for this to happen no coupling between
these fields is required, and hence we will assume $V = 0$ in the following discussion.
We take initial conditions at some initial time in which
both the scalar field $\phi$ and its Lee-Wick partner $\tphi$ are both oscillating about
their ground states, and that the positive energy density of $\phi$ exceeds the
absolute value of the negative energy density of $\tphi$, i.e. we start in a phase
dominated by regular matter. We assume that the universe is contracting with a
Hubble rate dictated by the Friedmann equations.

Initially both fields are oscillating and their energy densities both scale as $a^{-3}(t)$,
where $a(t)$ is the cosmic scale factor. Since $M \gg m$ while the energy
density of $\tphi$ is smaller than that of $\phi$, the amplitude $\tA$  of $\tphi$
must be much smaller than the amplitude $\cal{A}$ of $\phi$. During the initial
period of contraction, both amplitudes increase at the same rate. At some point,
however, $\tilde{A}$ becomes comparable to $m_{pl}$, the four dimensional
Planck mass. As we know from the dynamics of chaotic inflation  \cite{chaotic},
at super-Planckian field values $\phi$ will cease to oscillate - instead, it will enter
a ``slow-climb" regime, the time reverse of the inflationary slow-roll phase. During
this phase, the energy density of $\phi$ increases only slightly. However, $\tphi$
continues to oscillate and its energy density increases in amplitude exponentially
(still proportional to $a^{-3}$. The energy in $\tphi$ (i.e. its absolute value) 
will hence rapidly catch up with that of $\phi$. When this happens, $H$ will vanish.
Since the kinetic energy of $\tphi$ overwhelms that of $\phi$, $\dot{H} > 0$ and thus
a non-singular bounce will occur \cite{Cai2}, the universe and the universe will
begin to expand.

The matter bounce in the Lee-Wick scalar field model was analyzed in detail in
\cite{Cai2}. In particular, it was verified explicitly that initial vacuum fluctuations
on sub-Hubble scales in the contracting period develop into a scale-invariant
spectrum of curvature fluctuations on super-Hubble scales after the bounce.
A distinctive prediction of this scenario is the shape and amplitude of the
three-point function, the `bispectrum" \cite{bounceNG}.

However, the scalar field Lee-Wick bounce in unstable towards the addition of
radiation before the bounce
\footnote{The Lee-Wick matter bounce is also unstable against the addition
of anisotropic stress in the initial conditions. This is a well-known problem for
bouncing cosmologies which we will not further address in this paper.}: 
Since the energy density in radiation scales as
$a^{-4}$ it becomes more important than that of $\tphi$ as the universe
decreases in size, and will hence destabilize the bounce. Can the addition
of a Lee-Wick partner to regular radiation help restore the bounce? This is
the question we ask in this work.
We will follow the same type of reasoning as above, but for the case of radiation:
we now introduce a Lee-Wick gauge field, the partner of the standard one, which will
initially be dominant. We use the Lagrangian for a U(1) Lee-Wick gauge boson 
(see \cite{Grinstein:2007mp}) to which we add a coupling term between the normal 
and the Lee-Wick field in order to allow the energy to flow from one component to the 
other. Our goal is to see if we can get a bouncing universe using this setup.

\section{The Model}

We will consider the radiation sector of Lee-Wick quantum electrodynamics and
will start with a higher derivative Lagrangian  \cite{Grinstein:2007mp} for a 
$U(1)$ gauge field $A_{\mu}$ of the form:
\be
\label{F1}
L_{hd} \, = \, - \frac{1}{ 4} \hat{F}_{\mu \nu} \hat{F}^{\mu \nu} 
+ \frac{1}{2M_A^2} {\cal{D}}^{\mu}\hat{F}_{\mu \nu}
{\cal{D}}^{\lambda}\hat{F}_\lambda^\nu \, ,  
\ee
where $F_{\mu \nu}$ is the field strength tensor associated with $A_{\mu}$ and
${\cal{D}}$ denotes the covariant derivative. Note the sign difference in the second 
term compared to \cite{Grinstein:2007mp} : This will prevent the appearance of a 
tachyonic massive Lee-Wick (L-W) gauge boson. The mass $M_{A}$ corresponds
to the mass of the new physics in the model. To solve the Hierarchy Problem of
the Standard Model, this mass should be of the order of $1 {\rm TeV}$.

The higher derivative terms in the above Lagrangian lead to an extra propagating
mode. We can isolate it using the usual Lee-Wick construction by introducing a
new field $\ti{A_a}$ (the Lee-Wick partner) which depends on derivatives of
the original field and adjusting the gauge fields such that the kinetic
term of the Lagrangian becomes diagonal in $A_{\mu}$ and $\ti{A_{\mu}}$.
We find  that the propagator for the $\ti{A_a}$ field has pole at $p^2= M_A^2$ and 
has an opposite sign compared to the normal one. Thus, it is a ghost field (with
the associated problems of instability and non-unitarity mentioned in the
previous section). The Lagrangian becomes:
\be
\label{F2}
L \, = \, - \frac{1}{ 4} (F_{\mu \nu} F^{\mu \nu}  - \ti{F_{\mu \nu}} \ti{F^{\mu \nu}})
+ c F_{\mu \nu} \ti{F^{\mu \nu}} +\frac{M_A^2}{2}\ti{A_a}\ti{A^a} \, .
\ee
We have added a coupling term, with coupling constant c, in order to allow 
the energy density to be able to flow from the normal field to the Lee-Wick field.
Since the Lee-Wick sector is not observed in experiments today, we choose the 
two fields to be weakly coupled. In the case when the coupling constant is equal 
to zero, $M_A$ is the mass of the L-W gauge field.

Note that the $U(1)$ gauge invariance of electromagnetism is broken by the
addition of the Lee-Wick sector. In addition to the problem of ghosts, this is
another serious potential problem for the model which we are currently
investigating. Given that gauge invariance is violated, we need to justify
our choice of the coupling between the two fields. We have used gauge
invariance and power counting renormalizability to pick out the term
we have added to the Lagrangian in order to describe the coupling.
If the entire Lagrangian were gauge-invariant, this would clearly be the
correct procedure. In the presence of a symmetry breaking term which
is very small (for large values of $M_A$) we can use gauge invariance
of the low energy terms in the action to justify neglecting small symmetry
breaking coupling terms if we are interested in energy transfer between the
two fields which should be operational already at low energies.

As our initial conditions in a contracting universe, we imagine that the usual
radiation field dominates the energy-momentum tensor. This implies that
we must set the initial amplitude of $\ti{A_{\mu}}$ to be very small compared
to that of the regular gauge field $A_{\mu}$. In this case, then
if $M_A$ is large enough compared to the experimental energy scale,
we would not expect to see the ghost radiation field in experiments.

The energy-momentum tensor following from (\ref{F2}) is
 \be
\label{stress}
T_{\mu \nu} \, = \, - \frac{1}{ 4}g_{\mu \nu} (F_{\la \s} F^{\la \s} -\ti{F_{\la \s}} \ti{F^{\la \s}})
         +  { F_{\mu}}^ {\la} F_{\nu \la} 
         -   {\ti{F_{\mu}}^{ \la}} \ti{F_{\nu \la}}
         +  \frac{1}{2} g_{\mu \nu}{M_A}^2\ti{A_a}\ti{A^a}
          -  M_A^2  \ti{A_\mu}\ti{A_\nu} 
          -  4 c F_{\mu \la}\ti{F_\nu ^\la}
\ee
 and its trace is, contrary to the case of pure radiation, non-zero: 
 \be
 T^{\mu}_{\mu} \, = \, M_A^2   \ti{A_a}\ti{A^a} \, .
 \ee

The energy density is equal to:
\be
\label{stress00}
T_{00} \, = \, \frac{1}{ 4} (F^2 -\ti{F}^2) - c F_{\la \s} \ti{F}^{\la \s} +
                 {F_{0}}^{\la} F_{0 \la}  - {\ti{F_{0}}^{ \la}} \ti{F_{0 \la}}
                 - M_A^2  ( \frac{\ti{A}^2}{2}+\ti{A_0}^2) - 4 c F_{0 \la}\ti{F_0 ^\la} \, .
\ee
We can split this into three different terms, 
the contribution of normal radiation,
 \be
 \rho_A \, = \, \frac{1}{4} (F^2 +{F_{0}}^{\la} F_{0 \la}) \, , 
 \ee
 the contribution from Lee-Wick radiation,
 \be
  \rho_{\ti{A}} \, = \, -\frac{1}{ 4}(\ti{F}^2+{\ti{F_{0}}^{ \la}} \ti{F_{0 \la}}) 
      - M_A^2  ( \frac{\ti{A}^2}{2}+\ti{A_0}^2) \, ,
 \ee
and the term coming from the mixing between the two fields,
\be
\rho_{A-\ti{A}} \, = \, -c (F_{\la \s} \ti{F^{\la \s}} + 4  F_{0 \la}\ti{F_0 ^\la}) \, .
\ee

The equation of state is like that of radiation but with an additional term proportional to the
mass of the Lee-Wick gauge field:
\be
\label{w}
w \, \equiv \, \frac{p}{\rho} \, = \, \frac {\rho} {3\rho} + \frac {T^{\mu}_{\mu}} {3\rho} 
\, = \, \frac{1}{3} + \frac{M_A^2\ti{A_a}\ti{A^a}} {3T_{00}}
\ee
We note that this expression is valid only when the total energy density is non-zero, 
and thus it would not be valid at the bouncing point if there were a bounce.

We can actually define three different equation of state parameters, one for  each type of 
energy:
\ba
w_A \, &=& w_{A-\ti{A}} \, = \, {\frac{1}{3}} \,\,\,\,  {\rm{and}} \nonumber \\
w_{\ti{A}} \, &=& \, {\frac{1}{3}}+\frac{M_A^2\ti{A_a}\ti{A^a}} {\rho_{\ti{A}}} \, ,
\ea
the last of which is non-constant in time.
The equation of state parameter for the coupling term is the same as the one for 
normal radiation since the trace of the coupling energy-momentum tensor vanishes.

Our goal is to see under which conditions the above matter Lagrangian leads to a
cosmological bounce. We will initially turn off the coupling between the two fields
(i.e. set $c = 0$), derive the solutions of the equations of motion for both fields,
and study what scaling with the cosmological scale factor $a(t)$ these solutions
imply for the three contributions to the energy density discussed above. 
We find that - unlike what happens for the scalar field Lee-Wick model of
\cite{Cai2} - there is no mechanism which leads to a faster increase in the energy
density of the Lee-Wick partner field than that of the original radiation field. Thus,
a bounce can only occur if there is a mechanism which drains energy from the
original gauge field sector to the Lee-Wick partner field. It is for this reason that we
have introduced a direct coupling term between the two fields in our Lagrangian.
We will then study the effects of the coupling between the two fields, working
in Fourier space and making use
of the Green function method. We find that the sign of the energy transfer
depends not only on the sign of the coupling coefficient $c$, but also on the
phases of the oscillations of the two fields. Averaging over the phases, we find
no net energy transfer, and hence there can be no cosmological bounce.

As initial conditions we choose a state in the contracting phase in which the
regular radiation field is in thermal equilibrium at some initial time $t_i$. Since
we want to start with a state which looks like the time reflection of the state we
are currently in, we assume that the energy density in $\ti{A_{\mu}}$ is initially
sub-dominant. We, however, do assume that $\ti{A_{\mu}}$ has excitations for
modes with wave-number comparable to the initial temperature.

In the absence of coupling between the two fields, the distribution of $A_{\mu}$
would remain thermal, with a temperature $T$ which blue-shifts as the universe
contracts. The corresponding energy density would scale as $a^{-4}$. The
presence of coupling will lead to a departure from thermal equilibrium. We
will assume, however, that $a(t)$ continues to scale like $\sqrt {t}$, the scale 
factor of radiation. If there were a bounce, this approximation would fail at some point
sufficiently close to the bounce time.
       
\section{Equations of Motion}

The equations of motion obtained from varying the Lagrangian with respect to 
${A_\mu}$ and  $\ti{A_\mu}$ are:
\ba
\label{maxA}
\partial_{\mu} (F^{\mu \nu} - 2c\ti{F^{\mu \nu}}) + 3H ( F^{0 \nu}-2c\ti{F^{0 \nu}}) \, &=& \, 0 \\
-M_A^2\ti{A^{\nu}} + \partial_{\mu} (\ti{F^{\mu \nu}} + 2cF^{\mu \nu}) + 
3H (\ti{F^{0 \nu}}+2cF^{0 \nu}) \, &=& \, 0 \, .
\ea
Combining them, we find that  the L-W field will act as a source term for the normal field:
\be
\label{maxA2}
\partial_{\mu} F^{\mu \nu} + 3H F^{0 \nu} \, = \, \frac{2cM_A^2}{1+4c^2}\ti{A^{\nu}}
\ee
but that the L-W field is decoupled from the normal one and therefore 
only depends on the initial conditions:
\be
\label{maxALW2}
\partial_{\mu} \ti{F^{\mu \nu}} + 3H \ti{F^{0 \nu}}-\frac{M_A^2}{1 + 4c^2}\ti{A^{\nu}} \, = \, 0 \, .
\ee
From this last equation, we can also read off the new mass which the Lee-Wick partner field
obtains in the presence of coupling: ${M'_A}=\frac{M_A}{\sqrt{1+4c^2}}$,
which is about the same as $M_A$ at weak coupling. We can notice that at very strong coupling, 
the L-W gauge field becomes massless and therefore would evolve like a normal photon.

We will consider a homogeneous and isotropic universe with metric 
\be
ds^2 \, = \, - dt^2 + a^2(t) [dx^2 + dy^2 + dz^2] \, ,
\ee
where $t$ is physical time, $x, y$ and $z$ are the three spatial co-moving coordinates,
and we have for notational simplicity assumed that the universe is spatially flat.

Since the equations of motion are linear, we can work in Fourier space, i.e. with
plane wave solutions. There will be no coupling between different plane
waves. For simplicity, we focus on waves propagating along the z-axis with 
the same wave number, $k$, for the Lee-Wick and the normal gauge field. 
We work in the real basis of Fourier modes $cos(kz)$ and $\sin(kz)$.

Without loss of generality we can restrict attention to one polarization
mode which we take to be the electric field in the $x$ direction and the
magnetic field in the $y$ direction.
In this case, the only non-zero components of the field strength tensors are :
 $F^{0 1}, F^{1 3}\, \, \ti{F^{0 1}}$ and $\ti{F^{1 3}}$.
Using the temporal gauge where : $\ti{A_0}=A_0 =0$, we find that only the 
first component of the gauge fields are non-zero, and we can make the 
ansatz 
\be
A^1(k,t) \, = \, f(t)cos(kz) \,\,\,\, {\rm and} \,\,\,\, \ti{A}^1(k,t) \, = \, g(t)cos(kz)
\ee
or equivalently 
\be
A_1(k,t) \, = \, a(t)^2 f(t)cos(kz) \,\,\,\, {\rm and} \,\,\,\, \ti{A}_1(k,t) \, = \, a(t)^2g(t)cos(kz) \, .
\ee

From (\ref{maxA2}) and (\ref{maxALW2}),we obtain two linear second order differential equations with a damping term for the coefficient functions $f(t)$ and $g(t)$:
\ba
\label{f}
\ddot{f}(t) + H \dot{f}(t) + [\frac{k}{a(t)}]^2 f(t) \, &=& \, -\frac{2c}{1+4c^2}M_A^2 g(t) \\
\label{g}
\ddot{g}(t) + H \dot{g}(t) + [[\frac{k}{a(t)}]^2+\frac{M_A^2}{1+4c^2}] g(t) \, &=& \, 0 \, .
\ea
{F}or $\frac{k}{a} \ll \frac{M_A}{\sqrt{1+4c^2}} $, the L-W field behaves as a 
harmonic oscillator with angular frequency $\frac{M_A}{\sqrt{1+4c^2}}$. As
a consequence of the cosmological dynamics the oscillator undergoes
damping (in an expanding universe) or anti-damping (in the case of interest
to us, that of a contracting universe). The regular radiation field satisfies
the equation of a driven oscillator, again subject to cosmological damping
or anti-damping. Notice that(\ref {f}) has a particular solution  $f_{p}(t)=2 c g(t)$.
The driving term can lead to energy transfer between the regular radiation field
and its L-W partner. In the following we wish to study if the energy transfer is
able to drain enough energy from the regular radiation field to enable a bounce
to occur.

To solve these equations for any H(t), it is easier to use the conformal time  
$\eta=\int \frac{dt}{a}$ and to make things clearer, we introduce new functions 
$u$ and $v$ such that  $u(\eta) = f(\eta)$ and $v(\eta) = g(\eta)$.
Equations (\ref {f}) and (\ref {g}) can thus be rewritten as:
\ba
\label{f2}
u''(\eta) + k^2 u(\eta) \, &=& \, -a(t)^2\frac{2c}{1+4c^2}M_A^2  v(\eta)\\
\label{g2}
v''(\eta) + [k^2+{a(t)}^2\frac{M_A^2}{1+4c^2}] v(\eta) &=& \, 0 \, ,
\ea 
where $'$ denotes the derivative with respect to $\eta$.
For a radiation dominated universe, we have $a(\eta)=\eta$.

{F}rom (\ref{f2}) we see that in the absence of coupling we get 
simple oscillations in conformal  time with frequency $k$ for the normal 
gauge field. For the L-W field  we get  oscillations in conformal time, 
with a time dependant frequency $\ti{k}=\sqrt{k^2+{a(t)}^2\frac{M_A^2}{1+4c^2}}$ .
In physical time these correspond to:
\ba
\label{f3}f(t) \, &=& \, C cos(2 \sqrt{t} k + \phi) \\ 
g(t) \, &=& \, \frac{\alpha}{t^{\frac{1}{4}}}WhM(\frac{-i k^2 \sqrt{1+4 c^2}} {2M_A},\frac{1}{4},\frac{2 i M_A t}{\sqrt{1+4 c^2}})
+ \frac{\beta}{t^{\frac{1}{4}}}WhW(\frac{-i k^2 \sqrt{1+4 c^2}} {2M_A},\frac{1}{4},\frac{2 i M_A t}{\sqrt{1+4 c^2}})
\ea
where $WhM$ and $WhW$ are the Whittaker functions (see e.g. \cite{Whit}),
$\alpha$ and $\beta$ are constants characterizing the phase of $g(t)$ and $\phi$ is
the phase of $f(t)$.

Before discussing the solutions of these equations we must specify our initial conditions.
We consider a contracting phase dominated by regular radiation. Since we have
in mind an initial state which looks like the time reverse of a state in the early
radiation phase of our expanding cosmology, we will start at some time $t_i$ in thermal equilibrium
with a temperature much smaller than the Planck scale. The occupation numbers of
the Fourier modes of the regular radiation field are hence given by the thermal
distribution, with the peak wave-number being set by the temperature and hence
much larger than the Hubble rate. We are thus considering modes inside the Hubble radius.
Since we are interested in studying the possibility of obtaining a bounce, we will work
at temperature higher than the mass $M_A$.

We assume that the energy density of the L-W radiation field is sub-dominant at the
initial time $t_i$. The most conservative assumption is that the distribution of
wave-numbers is also peaked at the initial temperature. These assumptions
will allow us to pick out the limiting cases of the solutions of the above
equations (to be discussed in the following section) which are relevant for us.
   
\section{Solutions}
   
\subsection{Solutions for the Lee-Wick field}

Depending on whether the physical wave-number is larger or smaller than the mass of the 
L-W gauge field, $M_A'=\frac{M_A}{\sqrt{1+4c^2}}$, we get different behaviors for
the solution $g$. Since we are interested in exploring the solutions at high
densities, close to the hypothetical bounce point, we will assume that the
temperature is larger than the mass L-W field. We will focus on wave-numbers
close to the peak of the thermal distribution function, and hence $k/a > M_A'$.
In this limit, the solutions for the L-W gauge field will simply be oscillating  
in conformal time with frequency $k$:
\be
\label{g5}
g(t) \, = \, \ti{C} cos(\eta k) \, = \, \ti{C} cos(2 \sqrt{t} k) \, , 
\ee
where we have used the scaling of $a(t)$ of a radiation-dominated universe to express
the conformal time $\eta$ in terms of physical time $t$, and where $\ti{C}$ is a constant
amplitude.

The normal gauge field satisfies a harmonic oscillator equation with a driving term with
which the L-W field acts on it. The strength of the driving term is proportional to
the coupling constant $c$ in the Lagrangian. The general solution of the inhomogeneous
equation for $u$ is the general solution of the homogeneous equation plus a particular
solution of the inhomogeneous equation whose amplitude is proportional to $c$ and
which can be determined using the Green function method (see later).
The homogeneous solution for $u$ is oscillating with frequency $k$.

For large wavelength, i.e. $\frac {k}{\eta} \ll  M_A' $
the solutions for $g$ behave like a combination of  Bessel functions.
\be
\label{g4}
g(t) \, = \, \alpha \ t^{\frac{1}{4}} J(\frac{1}{4},\frac{M_A t}{\sqrt{1+4 c^2}})
+ \beta \ t^{\frac{1}{4}}Y(\frac{1}{4},\frac{M_A t}{\sqrt{1+4 c^2}}) \, ,
\ee
where $\alpha$ and $\beta$ are constants that can be determined using the initial conditions
and $J$ and $Y$ are, respectively, the Bessel functions of order $\frac{1}{4}$ of the first and 
the second kind.

A more physical way of understanding the behavior is to rewrite the solutions in the asymptotic 
limits. For large values of $t$ and for $ M_A' t \gg |\frac{1}{16}-1|$, the L-W gauge field oscillates 
with a frequency corresponding to the mass of the L-W gauge field, $M_A'$. Indeed, in this case :
\ba
\label{g4a}
     J(\frac{1}{4},M_A't) \, &\approx& \,  \sqrt{\frac{2}{\pi M_A't}} \cos \left(M_A' t-\frac{3\pi}{8}  \right)\\
     Y(\frac{1}{4},M_A't) \, &\approx& \,  \sqrt{\frac{2}{\pi M_A't}} \sin \left(M_A' t-\frac{3\pi}{8}  \right)
\ea
Therefore, in this limit the L-W gauge field scales like  $g(t)\propto  t^{-1/4} \sim a(t)^{-1/2}$ when we 
are in a radiation dominated period, which we are in the setup we are interested in where the
initial state is dominated by regular radiation.

To better understand the behavior of the solutions in the small $k$ limit and at 
large times, we can rewrite the solution using powers of the scale factor. The two independent
solutions are
\be
\label{g41}
g(t) \, \approx \, a(t)^{-\frac{1}{2}} e^{\pm \int{\frac{1}{2} \sqrt{H (t)^2 -\frac{4 M_A^2}{1+4c^2}}dt}} \, ,
\ee 
though this expression is valid only when the square root term in the exponential is 
approximately constant. Choosing the initial time $t_i$ such that 
$\frac{M_A^2}{1+4c^2} \gg  H (t_i)^2$ we see that this inequality stays valid only a finite 
period of time since $ H(t_i)$  increases with time in a radiation phase of a contracting
universe. We immediately get $g(t) \propto a(t)^{-1/2} cos( \frac{M_A}{\sqrt{1+4c^2}}t)$ 
which is in agreement with  the behavior we found using asymptotic values of the 
Bessel functions.

In the opposite case, when $t$ is close to $0$ (and we are still considering
large wave-numbers), the asymptotic forms of the Bessel functions of 
first and second kind scale as a power of $t$:
\ba
    t^{\frac{1}{4}}J(\frac{1}{4},\frac{M_A t} {\sqrt{1+4 c^2}}) \, &=& \,
    \frac{2^{\frac{5}{4}}(\frac{M_A} {\sqrt{1+4 c^2}})^{1/4} \Gamma({\frac{3}{4}} )\sqrt{t}}{\pi}+o(t^2) \\
   t^{\frac{1}{4}}Y(\frac{1}{4},\frac{M_A t}{\sqrt{1+4 c^2}}) \, &=& \, 
    \frac{-2^{\frac{3}{4}}} {(\frac{M_A}{\sqrt{1+4 c^2}})^{1/4} \Gamma({\frac{3}{4}})}
    + \frac{2^{\frac{3}{4}} (\frac{M_A}{\sqrt{1+4c^2}})^{1/4}\Gamma({\frac{3}{4}}) \sqrt{t} }{\pi}
    + \frac{1}{3}\frac{2^{\frac{3}{4}} M_A^2 t^2} {(\frac{M_A}{\sqrt{1+4c^2}})^{\frac{1}{4}} (1+4 c^2)\Gamma({\frac{3}{4}})} + {\cal{O}}(t^2) \, .
\ea
If we choose  the amplitude of the two Bessel functions to be equal and opposite in (\ref{g4}), 
we get a cancellation of the square root term in g(t) and thus the L-W gauge field scales as
$g(t)\approx  C_3 -C_4 t^2+o(t^2)$. In the general case we get
$g(t)\approx  C_3 +C_5 \sqrt{t} $ where $C_3$ and $C_5$ are constants.

Note that the closer we get to $t = 0$, less and less modes will satisfy the condition 
$k \ll |\eta| \frac{M_A}{\sqrt{1+4c^2}} $. Instead, they will evolve into the large wave-number 
regime discussed at the beginning of this subsection. They will oscillate and behave 
exactly as normal radiation. 

We note that since g(t) is just oscillating , its effect on the normal field will decrease with time
in a contracting phase  as the source will 
scale as $a(t)^2 \sim t$ in a radiation dominated era and time runs from $-\infty$ to $0$ in the
contracting phase.

\subsection{Scaling of the Energy Densities}

The energy densities for each type of radiation can be rewritten in terms of f, g and their derivatives for each mode k by averaging $<cos(kz)^2>$ over the z-direction:
\ba
\label{ra}
\rho_A(t,k) \, &=& \, \frac{1}{4 a^2} [(\frac{k}{a})^2 f(t)^2 +\dot{f}(t)^2]\\
\label{rat}
 \rho_{\ti{A}}(t,k) \, &=& \, -\frac{1}{4 a^2} [(\frac{k}{a})^2 +\frac{M_A^2}{2})g(t)^2 +\dot{g}(t)^2]\\
\label{rata}
\rho_{A-\ti{A}}(t,k) \, &=& \, -\frac{c}{a^2} [(\frac{k}{a})^2 f(t)g(t) +\dot{f}(t)\dot{g}(t)]  \, .
\ea
Rewriting this in term of conformal time, $\eta$, we get :
\ba
\label{era}
\rho_A(\eta,k) \, &=& \, \frac{1}{4 a(\eta)^4} [u'(\eta)^2+k^2 u(\eta)^2 ]\\
\label{era2}
 \rho_{\ti{A}}(\eta,k) \, &=& \, \frac{-1}{4 a(\eta)^4} [v'(\eta)^2+[k^2+ \frac{M_A^2}{2} a(\eta)^2]v(\eta)^2 ]\\
\label{era3}
\rho_{A-\ti{A}}(\eta,k) \, &=& \, \frac{-c}{a(\eta)^4} [u'(\eta)v'(\eta)+k^2 u(\eta)v(\eta)] \, .
\ea

In the absence of coupling between the two fields the solutions for $u$ correspond to undamped
oscillations. Hence, the energy density of the regular radiation field scales as $a^{-4}$ as
we know it must. The contribution of all short wavelength modes to the L-W energy density
also scales as $a^{-4}$ since for these modes $v$ is oscillating with constant amplitude. The
coefficient is negative as expected for a ghost field.
The third energy density, that due to interactions, also scales as $a^{-4}$ for short wavelengths.

The contribution of long wavelength modes to the energy density of the L-W field and to the
interaction energy density scale as $a^{-p}$ with a power $p$ which is smaller than $4$.
For large times, the power $p$ is $3$ in the energy density for the L-W field, i.e. a scaling
like that of non-relativistic matter. Close to $t = 0$ the power changes to $p = 2$. This can
be seen most clearly from (\ref{rat}) and from the scalings of $g(t)$ derived earlier.

Hence, we conclude that in the absence of coupling between the two fields (i.e. for $c = 0$),
the energy density in the regular radiation field will dominate throughout the contracting
phase if it initially dominates, and hence no cosmological bounce will occur.
In fact, for temperatures $T < M_A^{'}$, modes of $v$ with values of $k$ close to the peak
of the thermal distribution scale as matter. Hence, the ratio of the energy density in
the L-W field to the energy density in the regular radiation field decreases which renders
it even more difficult to obtain a bounce. Once $T > M_A^{'}$, the energy densities in
both fields scale as radiation.

\subsection{Solution for the Regular Radiation Field}

We now consider the evolution of the regular radiation field in the presence of a
non-vanishing coupling with the L-W radiation field. Our starting point is the
set of equations of motion (\ref{f2}) and (\ref{g2}). From (\ref{g2}) it follows that
the ghost field $v$ evolves independently. In turn, it influences the evolution of
the regular radiation field $u$ as a source term. We expect the coupling constant
$c$ to be small.

First, we show that the correction to the energy density in the presence of
non-vanishing coupling is very small, namely of order $c^2$.
We observe that if we turn on the coupling, the following is a solution of
(\ref{f2}):
\be
u(\eta)_{c\neq 0} \, = \, u(\eta)_{c=0}+2 c v(\eta) \, .
\ee
Inserting this into $\rho_A(k,\eta)$ (see (\ref{era})) yields
\be
\rho_{A \ c\neq 0} \, = \, \rho_{A \ c=0} - 4 c^2 (\rho_{\ti{A}} + \frac{M_A^2}{4} a(\eta)^{-2}v(\eta)^2 )
+ \frac{c}{a(\eta)^4} [u'(\eta)v'(\eta)+k^2 u(\eta)v(\eta)] 
\ee
Note that $\rho_{\ti{A}}$ and $v$ stay the same when we turn  the coupling on.
We also have a change in the expression for the coupling term in the energy density 
since it also depends on u:
\be
\rho_{A-\ti{A} \ c\neq 0} \, = \, - \frac{c}{a(\eta)^4} [u'(\eta)v'(\eta)+k^2 u(\eta)v(\eta)] 
- \frac{2c^2}{a(\eta)^4}[v'(\eta)^2+k^2 v(\eta)^2] \, .
\ee
The total energy density when the coupling is turned on is
 \be \label{total}
 \rho_{tot \ c\neq 0} \, = \, \rho_{A \ c\neq 0} + \rho_{\ti{A}} + \rho_{A-\ti{A}} \, = \, 
 \rho_{A \ c=0} + (1+4c^2)\rho_{\ti{A}} - c^2 M_A^2 a(\eta)^{-2} v(\eta)^2 
 \ee
This looks very much like the total energy we had before adding any coupling 
($\rho_{tot \ c=0} = \rho_{A \ c=0} + \rho_{A-\ti{A}}$) but with two correction terms of 
order $c^2$. Both correction terms appear to decrease the total energy
density (recall that $\rho_{ti{A}}$ is negative). The second correction term (the last
term in (\ref{total}), however, increases less fast in a contracting background than 
the other terms, and the first correction term corresponds to a small 
time-independent renormalization of the energy density in the L-W field. Thus,
it appears that if the energy density of the regular radiation field dominates
initially, then it will forever and no bounce will occur. In the following we
will confirm this conclusion by means of an analysis which compares
solutions with and without coupling with the same initial conditions. 

The evolution of $u$ in the presence of the coupling with $v$ can be determined using
the Green function method. The general solution $u(\eta)$ of (\ref{f2}) is the sum of
the solution $u_0(\eta)$ of the homogeneous equation which solves the same initial
conditions as $u$ and the particular solution $\delta u(\eta)$ with vanishing initial
conditions. The particular solution is given by
\be \label{GF}
\delta u(\eta) \, = \, u_1(\eta) \int_0^{\eta} d \eta^{'} \epsilon(\eta^{'}) u_2(\eta^{'}) s(\eta^{'})
    -   u_2(\eta) \int_0^{\eta} d \eta^{'} \epsilon(\eta^{'}) u_1(\eta^{'}) s(\eta^{'}) \, ,
\ee
where $u_1$ and $u_2$ are two independent solutions of the homogeneous equation,
$\epsilon(\eta)$ is the Wronskian
\be
\epsilon(\eta) \, = \, \bigl( u_1^{'} u_2 - u_s^{'} u_1 \bigr)^{-1} \, ,
\ee
and $s(\eta)$ is the source inhomogeneity
\be
s(\eta) \, = \, - a^2 \frac{2c}{1 + 4 c^2} M_A^2 v(\eta) \, .
\ee
In our case, the solutions of the homogeneous equation are $u_1(\eta) = cos(k \eta)$ and
$u_2(\eta) = sin(k \eta)$ and the Wronskian is $\epsilon(\eta) = - 1/k$.    

Since it is less hard to imagine a bounce once the energy densities in both fields
scale as radiation, and since to study the possibility of a bounce it is important
to investigate the dynamics at very high temperatures when the bulk of the Fourier
modes of both fields scale as radiation, we will consider in the following Fourier
modes for which $v$ is oscillating.

We will now show that the sign of the energy transfer between the two fields depends
on the relative phase between the oscillations of $u_0(\eta)$ and $v(\eta)$. We are
interested in conformal time scales long compared to the oscillation time $k^{-1}$ but
short compared to the cosmological time. Hence, we can approximate the scale factor
in (\ref{GF}) by a constant. A simple calculation then shows that if we choose 
phases for which $v(\eta) = v_0 sin(k \eta)$ and $u_0 = {\cal A} cos(k \eta)$ then
\be
u(\eta) \, \simeq \, \bigl( {\cal A} - \frac{2 c v_0}{1 + 4 c^2} M_A^2 \eta \bigr) cos ( k \eta) \, .
\ee
For a coupling constant $c > 0$ this choice of phase hence leads to draining of energy
density from the regular radiation field. On the other hand, the phase choice
$v(\eta) = v_0 cos(k \eta)$ and $u_0(\eta) = {\cal A} sin(k \eta)$ leads to
\be
u(\eta) \, \simeq \, \bigl( {\cal A} + \frac{2 c v_0}{1 + 4 c^2} M_A^2 \eta \bigr) sin ( k \eta) 
\ee
and hence to a relative increase in the energy density of the regular radiation field.

We need to consider the full phase space of Fourier modes. Even if we only consider
modes with fixed value of $k$ given by the peak of the thermal distribution, we must
sum over the different angles. Since there is no reason why the phases for different
Fourier modes should be the same, we must take the expectation value of the
energy transfer averaged over all possible choices of phases. This average 
obviously vanishes. Hence, we conclude that without unnatural fine tuning of
phases it is not possible to obtain the required draining of the energy density
from $u$ to $v$.

\section{Conclusions and Discussion}

If the scalar field sector of the Lee-Wick Standard Model is coupled to Einstein
gravity, then - in the absence of anisotropic stress - it is known that a
bouncing cosmology can be realized. Since the energy density in
radiation increases at a faster rate in a contracting universe compared to
that of non-relativistic matter, the cosmological bounce is unstable to
the addition of radiation to the initial conditions early in the contracting
phase. However, one may entertain the hope that the presence of
the ghost radiation which is present in the Lee-Wick model might allow
a bounce to occur in analogy to how the presence of ghost scalar field
matter is responsible for the bounce in the scalar field Lee-Wick model.

For a Lee-Wick radiation bounce to occur, either the energy density
of the ghost radiation would have to increase faster intrinsically than that
of regular radiation, or there would have to be a mechanism which
drains energy density from the regular radiation sector to the ghost
sector. We have shown that neither happens, unless the initial phases
of regular and ghost radiation are tuned in a very special way. Thus,
we have shown that in the Lee-Wick Standard Model, the presence of
radiation prevents a cosmological bounce from occurring.

The methods we have used in this paper could be applied to other
proposals to obtain a bouncing cosmology by modifying the
matter sector. Rather generically, one needs to worry whether
any given proposal is robust towards the addition of radiative matter.
The stability can be studied using the methods we have developed.
Whether a channel to effectively drain energy density from radiation
to ghost matter will exist may depend rather sensitively on the
specific model. Here, we have shown that in the Lee-Wick Standard
Model this does not happen. The same Green function method
could be used to study the energy transfer in other models.

Cosmologies in which the bounce is induced by extra
terms in the gravitational sector such as in the ``non-singular
universe construction" \cite{BMS}, the model of \cite{Biswas} or
the Horava-Lifshitz bounce \cite{HLbounce} are more likely to be
robust against the addition of matter. Specifically, the constructions
of \cite{BMS,Biswas} are based on theories which are asymptotically
free in the sense that at high curvatures the coupling of any kind of
matter to gravity goes to zero. This means that a bounce will not
be effected by adding radiative matter. In Horava-Lifshitz gravity,
there are higher spatial derivative gravitational terms which act
as ghost matter scaling as $a^{-4}$ and $a^{-6}$. The latter
are present if we go beyond the ``detailed balance case" and we
allow for spatial curvature. In this case, once again radiative
matter can be added without preventing a cosmological
bounce.

\begin{acknowledgments}

We would like to thank Wang Yi for useful discussions. This research is
supported in part by an NSERC Discovery Grant, by funds from the
Canada Research Chair Program, and by a Killam Research Fellowship to RB.

\end{acknowledgments}

\end{document}